\global\def\draftcontrol{0}

   \def\versionno{ phase transition}

\catcode`\@=11

\expandafter\ifx\csname draftcontrol\endcsname\relax\global\def\draftcontrol{0}
\fi

{\count255=\time\divide\count255 by 60
\xdef\hourmin{\number\count255}
\multiply\count255 by-60\advance\count255 by\time
\xdef\hourmin{\hourmin:\ifnum\count255<10 0\fi\the\count255}}
\def\draftdate{\number\month/\number\day/\number\year\ \ \ \hourmin }

\newcommand\makepapertitle{\par
  \begingroup
    \renewcommand\thefootnote{\@fnsymbol\c@footnote}%
    \def\@makefnmark{\rlap{\@textsuperscript{\normalfont\@thefnmark}}}%
    \long\def\@makefntext##1{\parindent 1em\noindent
            \hb@xt@1.8em{%
                \hss\@textsuperscript{\normalfont\@thefnmark}}##1}%
     \newpage
     \global\@topnum\z@   
     \@makepapertitle
     \thispagestyle{empty}\@thanks
  \endgroup
  \setcounter{footnote}{0}%
  \global\let\thanks\relax
  \global\let\makepapertitle\relax
  \global\let\@makepapertitle\relax
  \global\let\@thanks\@empty
  \global\let\@author\@empty
  \global\let\@date\@empty
  \global\let\@title\@empty
  \global\let\title\relax
  \global\let\author\relax
  \global\let\date\relax
  \global\let\and\relax
  \def\version{\let\version\@version\@gobble}
}
\def\@makepapertitle{%
  \newpage
   \ifnum\draftcontrol=1 {}
   \version\versionno
   \vskip 3em%
   \else
   \hfill\hbox to 3cm {\parbox{4cm}{\@pubnum}\hss}%
   \vskip 3em%
   \fi
   \begin{center}%
   \let \footnote \thanks
     {\LARGE {\@title}}%
     \vskip 1.5em%
     {\normalsize
       \lineskip .5em%
       \begin{tabular}[t]{c}%
         \@author
       \end{tabular}\par}%
     \vskip 1.5em%
     {\@bstract}%
     \end{center}%
     \vskip 1.5em
     \@date%
   \par
}

\gdef\@pubnum{}
\def\pubnum#1{%
  \gdef\@pubnum{#1}}

\gdef\@bstract{}
\def\Abstract#1{%
  \gdef\@bstract{%
   \parbox{\textwidth-0pc}{%
   \centerline{\bf Abstract}\penalty1000%
\kern.2cm%
\noindent
\renewcommand\baselinestretch{1.0}%
{#1}}}
}

\def\ps@paper{\let\@mkboth\@gobbletwo%
     \ifnum\draftcontrol=1
    \def\@oddfoot{\hbox to \textwidth{\tiny \versionno \hfil\tiny\draftdate}%
    \hskip -\textwidth \hbox to \textwidth{\hfil\rm\thepage\hfil}}%
     \else\def\@oddfoot{\hbox to \textwidth{\hfil\rm\thepage\hfil}}
     \fi
     \let\@evenfoot\@oddfoot
}

\def\body{\clearpage
          \pagestyle{paper}
    }

\def\@version#1{\ifnum\draftcontrol=1
\typeout{}\typeout{#1}\typeout{}
\vskip3mm\centerline{\hbox{\fbox{\normalsize{\tt DRAFT -- #1 -- }
                   {\draftdate}}}}\vskip3mm
\fi}
\let\version\@version
\long\def\eqlabel#1{\ifnum\draftcontrol=1
                    \tag@false  
                    \tag*{(\theequation) \hbox to -0.2cm{\hspace{0cm}\small{#1}\hss}}
                    \refstepcounter{equation}
                    \edef\@currentlabel{\theequation}
                    \ltx@label{#1}          
                    \else
                    \label{#1}
                    \fi
                    }
\let\st@bibitem\@bibitem
\let\st@lbibitem\@lbibitem
\ifnum\draftcontrol=1
  \def\@bibitem#1{%
    \st@bibitem{#1}\a@@label{#1}\ignorespaces}
  \def\@lbibitem[#1]#2{%
    \st@lbibitem[#1]{#2}\a@@label{#2}\ignorespaces}
  \def\a@@label#1{%
    \gdef\a@lab{\smash{\normalfont\small#1}}
    \ifvmode
      \if@inlabel
        \global\setbox\@labels\hbox{%
          \llap{\a@lab\let\a@lab\relax
                \kern\@totalleftmargin\kern\marginparsep}%
          \box\@labels}%
      \fi
    \fi}
\fi

\documentclass[12pt,letterpaper]{article}

\usepackage{amsmath,amssymb,array,calc,epsfig}

\ifnum\draftcontrol=1
\fi

\renewcommand\baselinestretch{1.25}
\renewcommand\section{\@startsection {section}{1}{\z@}%
                                   {-3.5ex \@plus -1ex \@minus -.2ex}%
                                   {2.3ex \@plus.2ex}%
                                   {\normalfont\large\bfseries}}
\renewcommand\subsection{\@startsection{subsection}{2}{\z@}%
                                   {-3.25ex\@plus -1ex \@minus -.2ex}%
                                   {1.5ex \@plus .2ex}%
                                   {\normalfont\normalsize\bfseries}}
\renewcommand\subsubsection{\@startsection{subsubsection}{3}{\z@}%
                                   {-3.25ex\@plus -1ex \@minus -.2ex}%
                                   {1.5ex \@plus .2ex}%
                                   {\normalfont\normalsize\it}}
\renewcommand\paragraph{\@startsection{paragraph}{4}{\z@}%
                                   {-3.25ex\@plus -1ex \@minus -.2ex}%
                                   {1.5ex \@plus .2ex}%
                                   {\normalfont\normalsize\bf}}

\numberwithin{equation}{section}

\def\revise#1       {\raisebox{-0em}{\rule{3pt}{1em}}%
                     \marginpar{\raisebox{.5em}{\vrule width3pt\
                     \vrule width0pt height 0pt depth0.5em
                     \hbox to 0cm{\hspace{0cm}{%
                     \parbox[t]{4em}{\raggedright\footnotesize{#1}}}\hss}}}}

\def\caln         {{\cal N}}
\def\calo         {{\cal O}}

\def\tr           {\mathop{\rm Tr}}

\def\sqr#1#2{{\vcenter{\vbox{\hrule height.#2pt
 \hbox{\vrule width.#2pt height#1pt \kern#1pt
 \vrule width.#2pt}\hrule height.#2pt}}}}

\def\a{\alpha}

\def\e{\epsilon}

\def\aa1{\phi}
\def\cc1{\psi}

\def\t{\tau}

\newcommand{\eq}{\begin{equation}}
\newcommand{\eqx}{\end{equation}}
\newcommand{\eqn}{\begin{eqnarray}}
\newcommand{\eqnx}{\end{eqnarray}}

\catcode`\@=12

\begin{document}


\title{\bf 
``Black Universe'' epoch  in String Cosmology}
\pubnum{UWO-TH-08/6
}


\author{
Alex Buchel$ ^{1,2}$ and  Lev Kofman$ ^{3}$\\[0.4cm]
\it $ ^1$Department of Applied Mathematics\\
\it University of Western Ontario\\
\it London, Ontario N6A 5B7, Canada\\
\it $ ^2$Perimeter Institute for Theoretical Physics\\
\it Waterloo, Ontario N2J 2W9, Canada\\
\it $ ^3$CITA, University of Toronto\\
\it 60 St. George st.\\
\it Totonto, Ontario M5S 3H8, Canada
}

\Abstract{
String theory  compactification involves manifolds 
with multiple warp factors.   For cosmological applications, we often 
introduce a short, high-energy inflationary throat, and a long,
low-energy Standard Model throat. It is assumed   that at the end of inflation,
 the excited Kaluza-Klein modes from the Inflationary throat tunnel to the SM throat
 and reheat Standard Model degrees of freedom, which are attached  to  probe brane(s).
 However, the huge hierarchy of energy scales
can result in a  highly dynamic transition of the throat geometry. 
 We point out that in such a cosmological scenario the Standard Model
 throat (together with SM brane) will be cloaked by a Schwarzschild horizon, produced by  the  Kaluza-Klein modes
tunneling from the short throat. The Black Brane formation is dual 
to the first  order chiral phase transition of the cascading gauge theory.
We calculate the critical energy density corresponding the formation of the the BH horizon in the long throat.
We discuss the duality between   "Black Universe" cosmology and an expanding universe driven by the 
hot gauge theory radiation. We address the new problem of the hierarchical multiple-throat scenarios:
  SM brane disappearance after the decay of the BH horizon.}

\makepapertitle

\body

\version\versionno

\section{Introduction: Warped geometry, Duality and Cosmology}

There are various high dimensional formulations of particle physics theories, 
such as  fundamental string theory or phenomenological high dimensional constructions.
 Yet, most often string theory cosmology is reduced
 to the conventional  $3+1$ dimensional 
 universe, perhaps    with some unconventional ingredients.
Low energy effective description   is obtained from the high dimensional theory by integration over
compact inner dimensions
\begin{equation}
\label{reduction}
S_4=\int d^4x \sqrt{-g} {\cal L}_4=\int d^4x \sqrt{-g}   \int d^6y \sqrt{G} {\cal L}_{10} \ .
\end{equation}

An example recipe for   string theory cosmology  is to start with KKLT \cite{kklt} (or large volume) compactification
in the type IIB theory, include multiple throats with conifolds (attached to the
bulk Calabi-Yau) to provide  the hierarchy of masses and couplings,
and to engineer inflation in this $4+6$ background geometry    with  branes or  moduli fields. 
In such a cosmological scenario at the end point of inflation, reheating,    energy  is supposed to stream   through
the labyrinth of the compact manifold to find, in one of its corners,   the light standard model particles which eventually
heat  and fill up the  universe \cite{Kofman:2005yz,Dufaux:2008br,tye2,csu}. For instance, one of the scenario is a
 warped brane inflation
which occurs in the short (inflationary) throat at the GUT energy scale  which contains an
anti-brane $\overline{D3}$ at the tip attracting the
  mobile brane $D3$ \cite{K2LM2T}.
Brane-antibrane annihilation terminates inflation and  energy  cascades into  KK modes associated with the
short throat. Suppose the SM particles are localized at the probe brane(s) in another, long SM throat of TeV  scale.
It is often assumed that KK modes of the short throat tunnel into the long throat and subsequently
transfer their energy into SM sector.
All the calculations can be done in ten  dimensional warped throat geometry (with the radial coordinate $y$ along of the throat)
\begin{equation}
\label{warped}
ds^2=H(y)^{-1/2}(-dt^2+d\vec x^2)+H(y)^{1/2}G_{ab}dy^ady^b\,, 
\end{equation}
 in the supergravity approximation 
and then reduced to the effective four dimensional picture, according to the prescription (\ref{reduction}).
For the warped throats, one can use well-studied Klebanov-Strassler \cite{ks} solution with the deformed conifold at the tip,
to estimate KK masses, tunneling and decay rates, etc.

The ten dimensional warped throat  supergravity  solution (like KS geometry)  has a dual four dimensional gravity-free
 gauge theory  description. 
Two or more throats attached to the bulk CY will be dual to two or more gauge  theories
\begin{equation}
\label{gauge}
{\cal L}(\phi_1, \phi_2)={\cal L}_1(\phi_1)+{\cal L}_2(\phi_2)+\Delta {\cal L}(\phi_1, \phi_2)\,,
\end{equation}
where $\Delta {\cal L}(\phi_1, \phi_2)=\sum_n \frac{{\cal O}_n(\phi_1, \phi_2)}{M^n}$ is
an interaction term due to the high dimensional operators, $\phi_1$, $\phi_2$ are the fields in two gauge theories.
Excitations of KK modes in supergravity correspond to the glueball excitations in the dual picture (\ref{gauge}),
while the tunneling of KK modes between the throats has a  dual description in terms of the 
high dimensional operator  $\Delta {\cal L}(\phi_1, \phi_2)$.

A specific feature  of the cosmological onset  is that the two throats containing interacting   excitations   is 
a highly dynamic system with the huge hierarchy   of the energy scales.
Indeed, each of the gauge theories ${\cal L}_{1}(\phi_1)$,  ${\cal L}_{2}(\phi_2)$ has its own critical energy $\epsilon_{c}$
(or critical temperature $T_{c}$) which separates the lower energy confinement phase from the higher energy 
deconfinement plasma phase. On the other hand, the phase transitions at $\epsilon_{c}$ correspond to the
emergence of the strong gravity regime in the supergravity dual, manifested in the appearance of a
black hole (black brane) horizon across the throat, heuristically similar to the AdS/BH solution 
\begin{equation}\label{adsbh}
ds^2=\frac{r^2}{R^2}\left(-\left(1-\frac{r_g^4}{r^4}\right)dt^2+d\vec x^2\right)+R^2\frac{dr^2}{r^2\left(1-{r_g^4}/{r^4}\right)} 
+R^2 d\Omega_5^2 \ .
\end{equation}
where $r$ is related to the radial direction  $y$ in (\ref{warped}).
 Therefore, to understand string cosmology scenario with two throats of the inflationary  and SM
 scales one has to go well beyond the simple picture of KK modes tunneling between throats and
to recall the higher dimensional description of the throat geometry which can contain the BH horizon.
Moreover, the time-dependent  cosmological scenario addresses to the string theory the questions
how the BH horizon in the throat  appears, evolves and eventually disappears, and what may be the interesting consequences of 
this higher dimensional epoch of the evolution of the universe. This is very different from the simple picture of KK modes
 tunneling between throats. Moreover, as  the BH horizon screens the tip of the SM throat (perhaps also
the $D3$ branes that live in the throat), the  $d y$ integration in Eq.~(\ref{reduction}) is not valid so that  we have to deal with
the higher dimensional theory. These are  the problems we will try to formulate and discuss in this paper.

Let us recall the story of the high temperature phase transition /BH horizon duality.
Although the motivation of  each  original study
 of such a duality was  different, we will look at those
from the perspective of our string theory cosmology  scenario.
 Phase transition/BH duality is deeply connected with the correspondence between
supergravity in the $AdS_5 \times S^5$  and CFT  at high temperature \cite{witten}.
In the context of supergravity  throat solutions and gauge theories,
gravity dual to the supersymmetric $SU(K)\times SU(K+M)$ gauge theory\footnote{Assuming unbroken chiral symmetry.} 
is the Klebanov-Tseytlin throat solution in
the type IIB theory with fluxes
\cite{kt} which  has  a singularity at the conifold.
In \cite{b} it was proposed that gauge theory with the restored chiral symmetry
above the critical temperature $T_{c}$ corresponds to the  Schwarzschild horizon 
which cloaks the naked singularity of the KT solution or tip of the throat in the KS solution.
While the original purpose in \cite{b} was to resolve the singularity of the KT solution,
its implications may  go well beyond this. Klebanov-Strassler solution with deformed conifold 
 resolves the singularity problem of KT solution at zero temperature. However, it is the BH horizon 
that provides a resolution of the KT singularity at high temperature.
Technically, it is a challenge to 
 find analytic BH throat solutions in ten dimensional type IIB theory with fluxes. 
Supergravity solution with the regular BH horizon in the KT throat was constructed in \cite{b2, ghkt,abk}.
Gauge theory with the broken chiral symmetry at low temperature $T \ll T_{c}$ is dual to the regular throat solution,
while gauge theory with restored chiral symmetry at $T \gg T_{c}$ is dual to the generalized throat solution 
with the regular BH horizon across the throat. Finite temperature gauge fields dynamics corresponds to the
first order phase transition at $T_{c}$. 

Originally, the  KS throat solution was constructed as a supergravity dual to the cascading $SU(K)\times SU(K+M)$ gauge theory.
However, because this self-consistent geometry describes the throat with the regular tip at the finite distance, and because this
throat can be smoothly embedded in the CY manifold \cite{aha,kl}, KS solution is often evoked by cosmologists
as the model of the throat attached to the compact space. 

Correspondence between BH horizon and the phase transition was also discussed in the very different context, namely, in the
 phenomenological five-dimensional  Randall-Sundrum  model.
While usually the four-dimensional effective theory at the brane is derived from the bulk+branes system in the 
spirit of Eq.~(\ref{reduction}),
there was a proposal in \cite{apr} to extrapolate the ideas of holography 
for the RS braneworld and particle phenomenology on the brane.
Original high-temperature plasma phase was proposed to be dual to the bulk AdS/BH horizon in five dimensions, while 
 EW  phase transition at the brane  should  be dual to the disappearance of the horizon \cite{cnr}.

In the following we recall  the cascading gauge theory side of the duality.  After that we   explicitly evaluate  the
critical energy density $\epsilon_c$ for a given throat. This will allow us to lay down the new cosmological
scenario which contains higher dimensional stage of formation and later disappearance of the higher dimensional BH horizon.
Similarly, this dynamics is  relevant for the theoretical ``thought experiment'', if one is not interested in cosmology. 
Finally we discuss potential cosmological problems and consequences of this cosmological model and address interesting
 theoretical issues.

\section{Gauge Theory Perspective}
The gauge theory dual to a local warped geometry is a so called  ``cascading gauge theory''
introduced in \cite{kn,kt,ks}. This cascading gauge theory can be thought of as a specific $SU(K)\times SU(K+M)$
$\caln=1$ supersymmetric gauge theory, with a number of colors $K$ which runs logarithmically with the energy 
scale $\mu$ \cite{ks,b,k,b3}
\begin{equation}
K=K(\mu)\sim M^2\ \ln\left(\mu/\Lambda\right)\, \ ,
\eqlabel{nrun}
\end{equation}
where $\Lambda$ is the strong coupling scale of the cascading gauge theory.
Despite having an infinite number of degrees of freedom in the ultraviolet, cascading gauge theory is 
holographically\footnote{In the UV the cascading gauge theory 't Hooft coupling 
becomes strong and thus renormalization of this
gauge theory must be addressed in the framework of the dual gravitational description.}  
renormalizable as a four dimensional quantum field theory \cite{aby}.
At a given scale $\mu$ the gauge theory has two chiral superfields $A_1$, $A_2$ in the $(K+M,\overline{K})$
representation, and two fields $B_1$, $B_2$ in the $(\overline{K+M},K)$ representation. The superpotential of the model 
is 
\begin{equation}
W\sim \tr \left(A_i B_j A_k B_\ell\right) \e^{ik}\e^{j\ell}\,.
\end{equation}
The two gauge group factors have gauge couplings $g_1$ and $g_2$. Under the renormalization group flow 
the sum of the coupling does not run
\begin{equation}
\frac{4\pi}{g_1^2}+\frac{4\pi}{g_2^2}={\rm constant} \,,
\eqlabel{gsum}
\end{equation}
while the difference is 
\begin{equation}
\frac{4\pi}{g_2^2}-\frac{4\pi}{g_1^2}\sim M\ \ln\left(\mu/\Lambda\right)\left[3+2(1-\gamma)\right] \,,
\eqlabel{gdiff}
\end{equation}
where $\gamma$ is the anomalous dimension of operators $\tr A_i B_j$. It is clear from \eqref{gsum}, 
\eqref{gdiff} that starting at some energy scale and flowing either to the UV or the IR one inevitably 
encounters a Landau pole:  one of the two gauge couplings will become infinitely large. In \cite{ks}
it was argued that extension of the RG flow past the infinite couplings is achieved by a cascade of 
self-similar Seiberg duality \cite{s} transformations on the strongly coupled gauge group factor. 
At each duality step $K\to K+M$ for the RG flow to the UV, and $K\to K-M$ for the RG flow to the IR,
leading to the effective logarithmic running of the number of colors \eqref{nrun}. 

Cascading gauge theory confines in the IR; it has a classical $U(1)_R$ symmetry which is explicitly 
broken to $\mathbf{Z}_{2M}$ 
by the anomaly, and then it is broken spontaneously by the gluino condensate to $\mathbf{Z}_2$ \cite{ks}. 
In \cite{b} it was pointed out  that at sufficiently high temperature, the R-symmetry of the gauge theory
is restored; moreover, this restoration is accompanied by a first order deconfinement phase transition\footnote{The high
temperature deconfinement state of the cascading gauge theory was studied in \cite{b,b2,ghkt}.}.    
Such a phase transition was recently identified in \cite{abk}. A summary of the proposal \cite{b} and the 
detailed analysis \cite{abk} is that a cascading gauge theory at equilibrium has a critical energy density,
which we refer to as $\e_{c}$, such that for density $\e>\e_{c}$ it is in a deconfined 
phase with $\calo \left(K(\e)^2\right)$  entropy.

Several comments are in order before we finish the gauge theory discussion.

 First, the deconfinement phase transition at $\e> \e_{c}$ assumes that the boundary gauge theory 
is in flat space-time, or at least $\e_{c}\gg R_4^{2}$. Here,   $R_4$ is the Ricci scalar of the background metric. 

 One expects a chiral symmetry restoration ( but {\it without} the deconfinement ) phase transition in curved 
space-time. For a cascading gauge theory on $S^3$ this was demonstrated in \cite{bt} while for the 
cascading gauge theory in $dS_4$ this was discussed in \cite{b4}. In both cases the background curvature serves 
as a regulator that cuts off the IR physics associated with the gluino condensate. By studying  a  $D3$-brane probe 
in the deformed geometry\footnote{In the $dS_4$ case this was done in \cite{br}.} it is easy to see that 
the spectrum of fluctuations in the deformed throat is gapped.  

 In principle, for a cascading gauge theory in flat space-time one could imagine two separate phase transitions:
a confinement/deconfinement one, and the chiral symmetry restoration transition. For the  scenario discussed in this paper 
it is important whether or not there is a deconfined phase with a broken chiral symmetry. Such a phase, were it to exist, 
is expected to have a critical energy density higher than $\e_{c}$. A detailed analysis \cite{ab} indicates that such a phase 
is not realized. In other words: the hot cascading gauge theory plasma cools though a first order phase transition 
where the  chiral symmetry is broken only  when the theory confines.

\section{Critical Energy Density in the  Throat}

In this Section we turn to the  throat geometry. We will 
   compute the critical energy density for weakly curved 3+1 dimensional space-time 
(modeling our universe) that is necessary to hide it behind the horizon in the ambient warped throat geometry. 
Each throat has its own critical density, which corresponds to  horizon formation, but for our scenario we will focus on  
 $\epsilon_c$ for the SM throat.

We assume that our universe (in the low energy limit where the  energy density in all throats
is  much lower than $\epsilon_c$) is part of the KS  geometry
which away from  the tip is locally  described by the KT geometry 
\begin{equation}
ds_{10}^2=H^{-1/2}(r)\ \eta_{\mu\nu}dx^{\mu}dx^{\nu}+H^{1/2}(r)\left(dr^2+r^2 d s_{T^{1,1}}^2\right)\,,
\eqlabel{kt} 
\end{equation}
where 
\begin{equation}
H(r)=\frac{R_+^4+R_-^4\ \ln\left(\frac{r}{R_+}\right)}{r^4}\,,
\eqlabel{warp}
\end{equation}
with 
\begin{equation}
R_+^4=\frac{27\pi }{4}\a'^2 g_s M K\,,\qquad R_-^4=\frac{81}{8}\a'^2g_s^2 M^2\, \ ,
\eqlabel{rprm}
\end{equation}
and $M$ and $K$ are the integer numbers associated with the quantization conditions of the form fields 
  fluxes of the type IIB theory
which generate the throat geometry (\ref{kt}). 
The  local geometry \eqref{kt} is holographically dual to the four-dimensional  cascading gauge theory reviewed in 
the previous section. 
For large enough energy density this  cascading gauge theory undergoes a deconfining phase transition \cite{b,abk},
which is reflected in the formation of a Schwarzschild horizon\footnote{As we will explain
in the following Sections,  in our scenario this energy density is accumulated 
from the ``gravitational collapse'' of the excited KK modes initially produced from  the $D3\overline{D3}$ annihilation in the 
inflationary throat.} in \eqref{kt}. The type IIB supergravity background corresponding to a Schwarzschild horizon
in asymptotic geometry \eqref{kt} takes form \cite{abk}\footnote{The frames $\{e_{\theta_a},e_{\phi_a}\}$ are defined as
in \cite{aby}, such that the metric on a unit size $T^{1,1}$ is
given by $\left(e_\psi^2\right)+ \sum_{a=1}^2
\left(e_{\theta_a}^2+e_{\phi_a}^2\right)$.}:
\begin{equation}
\begin{split}
ds_{10}^2=&h^{-1/2}(2Y-Y^2)^{-1/2}\left(-(1-Y)^2 dt^2+d \vec x^2\right)+G_{YY}dY^2\\
&+h^{1/2} [f_2\ \left(e_\psi^2\right)+ f_3\ \sum_{a=1}^2
\left(e_{\theta_a}^2+e_{\phi_a}^2\right)]\,,
\end{split}
\eqlabel{ktm}
\end{equation}
where $h$, $f_2$ and $f_3$ are some functions of the radial
coordinate $Y\in [0,1]$.
There is also a dilaton, and form fields ( see \cite{abk} for details).
Notice that the radial coordinate in \eqref{ktm} is gauged-fixed\footnote{Because we gauge-fixed 
the radial coordinate, there is a constraint equation coming from the equation of motion of this variable; 
this equation can be solved to determine $G_{xx}$ \cite{abk}.} so that
\begin{equation}
 \eqlabel{xfixing} \frac{G_{tt}}{G_{ii}} = -(1-Y)^2\, \ . 
\end{equation}
The new radial coordinate $Y$ is related asymptotically to  $r$ in \eqref{kt} as follows
\begin{equation}
Y\sim \frac{1}{r^4}\,,\qquad r\to \infty\,.
\eqlabel{xrrel}
\end{equation} 
Type IIB supergravity equations of motion in the background metric \eqref{ktm} are solved, subject 
to the following boundary conditions:\\
i) near the boundary ($Y\to 0$ or $r\to \infty$) the black hole metric \eqref{ktm} approaches 
the Klebanov-Tseytlin geometry \eqref{kt};\\
ii) the hypersurface $Y=1$ is a regular Schwarzschild horizon of the metric \eqref{ktm}; the latter is equivalent to requiring that  
all the warp factors $h$, $f_2$ and $f_3$ 
are positive at $Y=1$.

When the temperature $T$ of the black hole \eqref{ktm} is much larger than the characteristic 
KK scale $m_{KK}$  deep inside the throat geometry \eqref{ktm}\footnote{See Eq.\eqref{KK} of the next Section.} 
\begin{equation}
T\gg \Lambda \sim m_{KK}\sim \frac{1}{R_-}\ e^{-\frac{2\pi K}{3Mg_s}}\,,
\eqlabel{hight}
\end{equation}
the black hole geometry and its thermodynamics can be determined analytically \cite{ghkt,aby,abk}.
Specifically, we find for $T\gg \Lambda$ 
\begin{equation}
f\simeq -\frac{1}{8}\pi^2 K(T)^2\ T^4\sim -M^4 T^4 \left(\ln \frac{T}{\Lambda}\right)^2 \,,\qquad \e\simeq -3 f\,,
\eqlabel{fe}
\end{equation} 
for the black hole free energy density $f$ and  the energy density $\e$. 
We have used \eqref{nrun} to arrive at \eqref{fe}. Here $K(T)$ is the number of degrees of freedom, c.f. Eq.~(\ref{nrun}). 

For temperature $T$ of order $\Lambda$ the black hole thermodynamics can be studied only numerically. 
It was found in \cite{abk} that there is a critical temperature, corresponding to a critical energy 
density $\e_c$, such that the free energy density of the black hole \eqref{ktm} vanishes precisely 
at $\e=\e_c$, and becomes positive for $\e<\e_c$. 
 Thus we expect 
a first order deconfinement/confinement phase transition in the cascading gauge theory plasma once its energy 
density becomes  less than $\e_c$.
This first  order phase transition (which occurs via bubble nucleation and percolation) shall be  dual to the
inhomogeneous melting of the BH horizon (which before this is translationally  invariant in the $\vec x$ directions).

 From the point of view of the usual $3+1$ dimensional GR, the  disappearance of the horizon is  rather unusual.
 However, the disappearance  of a higher dimensional horizon is a familiar phenomenon, and we shall comment on it.
 A crucial observation is  that at the supergravity side, 
there is another geometry, besides the black hole geometry \eqref{ktm}, which also asymptotes to
 (a Euclidean version) of \eqref{kt}:
the KS geometry with a Euclidean time direction identified with a period of $1/T$. The latter solution has no horizon, 
it has a zero free energy, and is dual to a thermal gas of the confined cascading gauge theory.
The coexistence of these two solutions provides the possibility for   transitions between horizon and no-horizon geometrical phases
(through  instantons, see.e.g. \cite{cnr}).
The supergravity dual of the phase transition bears similarity  to the  Hawking-Page 
transition, which was identified by Witten as a supergravity dual to a (kinematic) confinement/deconfinement
transition of the $\caln=4$ SYM plasma on a three-sphere \cite{witten}. There is a notable difference however: 
while the Hawking-Page-Witten transition occurs in a finite volume\footnote{The transition disappears in the infinite volume.}, 
the phase transition in the cascading gauge theory occurs in the infinite volume. Thus, we expect it to proceed via the 
nucleation of bubbles of the stable phase, which further expand and 'remove' the horizon.     

Next, we evaluate $\e_c$ quantitatively.
In the notation of \cite{abk}, a  Schwarzschild horizon is formed in \eqref{kt} at 
 the critical energy density (at the gauge theory side)  which is given by equation (see also Eq.~(4.10) of \cite{abk})
\begin{equation}
\e_{c}=\frac{1}{4\pi G_5}\ a_0^2\,,
\eqlabel{ec}
\end{equation}
where the 5-dimensional Newton's constant $G_5$ is 
\begin{equation}
\frac{1}{G_5}=\frac{vol_{T^{1,1}}}{G_{10}}=\frac{16\pi^3}{27}\ \times \frac{1}{8\pi^6 g_s^2 \a'^4} \,,
\eqlabel{g5}
\end{equation}
and $a_0$ will be evaluated momentarily.
A careful matching of the asymptotic black hole geometry in \cite{abk} with \eqref{kt} leads to 
identification
\begin{equation}
\left(\frac{R_+}{R_-}\right)^4-\frac 14 \ln\frac{2 e R_+^4}{ a_0^2}=\frac 12 k_{c} \,,
\eqlabel{map}
\end{equation}
where 
\begin{equation}
k_{c}=0.25712(1) \,,
\eqlabel{kc}
\end{equation}
(see Eq.~(5.12) of \cite{abk}).

From \eqref{rprm}-\eqref{map} we conclude 
\begin{equation}
\e_{c}=2 M K e^{1+2 k_{c}}\, T_3\, e^{-4 A} \,,
\eqlabel{ec2}
\end{equation}
where $T_3$ is a $D3$ brane tension $T_3=\frac{1}{(2\pi)^3 g_s\a'^2}$
and $e^{-A}=e^{-\frac{2\pi K}{3Mg_s}}$
is the hierarchy warp factor of the throat \cite{gkp}. 
The four-dimensional energy-density of KK modes (located at the tip of the throat) is identified with  the
energy-density of the glueballs in the dual picture. 
Also notice that the bigger energy $\epsilon$ corresponds to the bigger portion of the throat cloacked by the horizon.

\section{Theory of  Multiple-throat Tunneling Revisited}

\begin{figure}[h]
\centerline{
\includegraphics[width=0.8\textwidth]{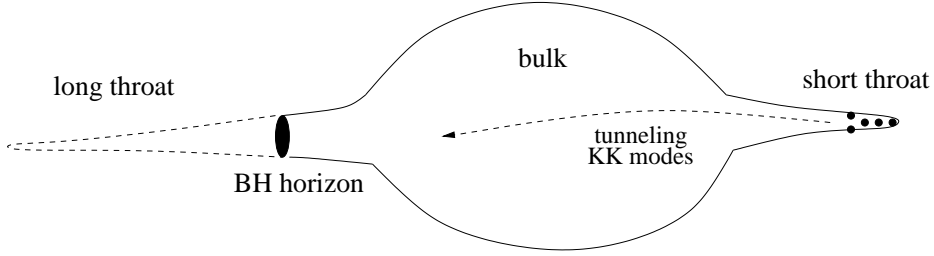}
}
\caption{Sketch of the compact manifold with two throats. }
\label{throats}
\end{figure}

Suppose compact manifold contains throats of significantly different warpings
which are attached to the bulk, as sketched in the Figure~\ref{throats}.
Suppose the short throat  is associated with the higher energy scale and is populated with  Kaluza-Klein excitations.
It is well known that the typical KK masses are of the order of
\begin{equation}\eqlabel{KK}
m_{KK}\simeq \frac{e^{-A}}{R} \ ,
\end{equation}
where $e^{-A}=H(y)^{-1/4}$ in the warping factor at the tip of the throat, $R$  is the
radius scale of the angular coordinates at the tip, for the metric (\ref{kt}) it is $R_-$.
Wave functions of the KK modes in the warped geometry are exponentially peaked around the tip of the throat,
as illustrated in the  Figure~\ref{throats}.

Let us set up the thought experiment with the static geometry of the  Figure~\ref{throats}, with the flat
outer space.  Suppose  the short throat is filled up with the
KK excitations of the mass \eqref{KK} and four-dimensional energy density $\epsilon_1$. For definiteness we can take
$e^{-A_1}=10^{-3}$, $R_1\simeq  10 \sqrt{\alpha'}\simeq 10^2 M_p^{-1}$ so that $m_{1KK}\sim 10^{13}$ GeV
 (index 1 attributes parameters to the short throat, while index 2 to the long throat).
Suppose $\epsilon_1 \sim T_3 \sim (10^{16}GeV)^4$. KK modes from the short throat
begin to tunnel into the long throat. Putting aside interesting subtleties (see e.g. \cite{Kofman:2005yz,tye2,csu,Heb})
let us use the estimation of \cite{Dimopoulos:2001ui} for the inter-throat tunneling rate
\begin{equation}\eqlabel{rate}
\Gamma_{tun} \simeq (m_{1KK}R_1)^4 e^{-A_1} \sim \frac{e^{-5A_1}}{R_1} \,,
\end{equation}
which gives us $\t_{tun}=1/\Gamma_{tun} \sim 10^{-26}$ sec. 
During this time the long throat is filling up with  KK excitations, each with  mass much lighter than $m_{1KK}$.
If we choose $e^{-A_2}=10^{-15}$, $R_2=R_1$, then $m_{2KK}\sim 10$ GeV.
Those modes again are accumulating\footnote{We can neglect the tunneling of KK modes from the long (SM) throat back
to the short (inflationary) throat as the latter is suppressed 
compare to \eqref{rate} by a factor  $\propto e^{5(A_2-A_1)}\propto 10^{60}$.} 
around the tip of the long throat, with ever increasing four-dimensional energy density $\epsilon_2(t)$.
However, gravitational backreaction of KK modes on the geometry of the long-throat becomes significant as soon  as $\epsilon_2(t)$
is approaching from below  the critical density $\epsilon_{2c}$ of the long throat.
Let us estimate $\epsilon_{2c}$ from Eq.~(\ref{ec2}).
For example, choosing $MK \sim 10^4$,    from   $\e_{2c}\sim \left( (MK)^{1/4} 10 GeV \right)^4$ we get 
   $\e_{2c}\sim (100 GeV)^4$, i.e. the scale of the 
EW phase transition.
 This is  much lower than 
the original energy density $\epsilon_1$ of the excitations in the system. It means that long before the time  $\t_{tun}$
 KK modes accumulating in the long throat  will completely change the
geometry of that throat. Actually, it happens instantly
 \footnote{This follows from the energy balance equation $\epsilon_2=\e_1 \left(1-e^{-t\Gamma_{tun}} \right)$.}.

 While it is immensely difficult  to follow the detailed complicated metamorphose of the compact 
space in the self-consistent supergravity formalism, the dual picture in the gauge theory side suggests that increasing
 energy of the plasma above $\e_{2c}$ corresponds to the formation of  the BH horizon in the throat geometry.
Since the chiral phase transition in the gauge theory plasma is a first order one, the formation of the BH horizon will be
dual to the formation of the bubbles of the new phases which eventually percolate.

Therefore we conjecture the following picture of the throat BH  formation. The horizon begins to cloak the tip of the throat
and propagates,  further screening a bigger and bigger part of the throat. This may occur in the manner  of  the
Choptuik critical collapse \cite{ch}. 
Formation of the horizon may occur not uniformly in space,  but  in  patches
which later percolate into the uniform horizon (translationary-invariant in  three-dimensional outer space).
 As KK modes from the short throat continue to
tunnel into the long throat, BH there absorbs all of them and it happens much faster that the tunneling time $1/\Gamma_{tun}$
(this timing is defined by the cross-section of the high dimensional black hole). 

All together, in our thought experiment we start with the excitations of the KK modes in the short throat and end up with the
high dimensional Black Hole cloaking the tip of the  long throat. From the four dimensional perspective, there are no 
 particles, but uniformly distributed energy density from the high dimensional BH. Again, the dual picture suggest 
this form of the energy density has the radiation dominated equation of state but without radiation!
Recall similar situation in the RS braneworlds with the AdS/Schwarzschild bulk geometry where four-dimensional
``dark radiation'' is associated with the projection of the five-dimensional Weyl tensor \cite{misao}.

\section{``Black Universe'' Cosmological Scenario}

 The setting of the thought experiment of the  previous Section takes place  naturally in the popular
string theory cosmological model. Indeed, in this model,  the warped brane inflation is based on the brane-antibrane 
interaction in the short (inflationary) throat,
 which provides very shallow effective four-dimensional  inflationary potential \cite{K2LM2T}.
 Four dimensional energy density of the branes is $\epsilon_1=2 T_3$.
At the end of inflation  the brane-antibrane pair annihilates and releases  energy into  KK modes excitations, e.g.
\cite{Barnaby,Kofman:2005yz,Frey}. KK modes from the short throat begin to tunnel into the long throat.
The results of  the previous section (where four dimensional outer space is not expanding)
indicate that  the cosmological scenario with the multiple throat geometry
of the inner manifold drastically differs from what was considered earlier in the literature on
warped brane string theory cosmological model.

In fact, expansion of the universe makes the multiple-throat cosmological scenario even more involved,
and in some aspects different from the  scenario without cosmological expansion considered of the previous Section.

During  $D3\overline{D3}$ branes inflation in the short throat  the Hubble parameter is
 $H = \sqrt{\frac{2T_3}{3M_p^2}} \sim 10^{13}$GeV. After inflation this value decreases with time as $1/t$.
Thus, initially expansion of the universe is significant and would dilute the energy density of
KK modes in the long throat, which are much lighter than that of the short throat ($m_{2KK} \ll m_{1KK}$). KK modes of the
long throat  behave as  radiation, while KK modes in the short throat have matter equation of state.
As a result  the tunneling will happen only  after the Hubble rate drops below the value equal to the 
tunneling rate \cite{csu}, $H\sim \Gamma_{tun} \sim 10$GeV.
Energy density in the long throat at this moment is
\begin{equation}
\e_2\sim M_{p}^2 H^2_{tun}\sim M_p^2 \Gamma_{tun}^2 \sim M_p^2 m^2_{1KK} e^{-8A_1} \ ,
\eqlabel{e2}
\end{equation}
in our example $\e_2\sim (10^9 GeV)^4$.

However, large value of the Hubble parameter generates mass gap of the KK modes $\Delta m^2_{KK}=2H^2$. For instance, it is  known
that massive gravitons in the four-dimensional de Sitter geometry   have the mass gap $2H^2$ \cite{higuchi}.
As long  as $H$ is large, KK modes of the long throat are not light, but become lighter and lighter as $H$ decreases.
As $H$ drops below $\Gamma_{tun}$, those KK modes can be treated as radiation and we return to the same
estimate (\ref{e2}) \footnote{Above estimate is correct provided that the problem of the angular KK modes \cite{Dufaux:2008br} is resolved  and 
 the decay of KK modes into
 gravitons is suppressed compare to  the 
tunneling time, see  \cite{csu} for details. Our choice of parameters respect  these conditions}.

Comparing  \eqref{e2} and \eqref{ec2} we find
\begin{equation}
\frac{\e_2}{\e_{2c}}\simeq\ \frac{M_p^2 m_{1KK}^2}{T_3}\ \frac{\left(e^{-A_1}\right)^8}{\left(e^{-A_2}\right)^4}\
\frac{1}{2 e^{1+2k_c} MK} \ , 
\eqlabel{e2e2c}
\end{equation}
which is $10^{28}$ in our example. 
Thus tunneling KK modes from the short throat collapse into a black hole in the long throat, see Figure 1. 
Our universe enters the {\it Black Universe} (BU) phase of its cosmological evolution. 
Because of the horizon, dimensional reduction (\ref{reduction}) is not relevant. Meanwhile,  high dimensional strong gravity theory
is very complicated.

Yet, the further evolution of the  Black Universe  phase can be  understood with the magic of duality. 
Had the long throat cloaked with the horizon been infinitely long, it would have been  holographically dual to 
the deconfined Klebanov-Tseytlin plasma (discussed in Section 2) in Minkowski space-time  without gravity.
In this case,  however,  the four dimensional Planck mass $M_p$ would be infinitely large. 
Gluing the long throat to a compact manifold produces a finite $M_p$, and in the dual picture this corresponds to the 
coupling of the  hot KT plasma to 4d gravity \cite{gr1,gr2,apr}.
In other words,  the compact manifold with strong gravity in two throats
is dual to the four dimensional theory (\ref{gauge}) where we will add the four-dimensional gravity $R_4$.
 Energy density of the deconfined KT plasma
will drive adiabatic expansion of the background space-time, while redshifting itself as radiation. 
Such an expansion continues until the plasma energy density redshifts to the critical energy density of the 
first order confinement transition \eqref{ec2}.
The Hubble scale at the beginning of radiation-dominated expansion is 
\begin{equation}
H_{initial}^{BU}\sim \sqrt{\frac{\e_2}{M_p^2}}\sim m_{1KK}e^{-4A_1} \,,
\eqlabel{hb}
\end{equation}
while the Hubble scale at the confinement transition is
\begin{equation}
H_{final}^{BU}\sim \sqrt{\frac{\e_{2c}}{M_p^2}}\propto 10^{-14}\ H_{initial}^{BU}\,.
\eqlabel{he}
\end{equation}
Since 
\begin{equation}
\frac{H_{initial}^{BU}}{\e_2^{1/4}}\sim \sqrt{\frac{m_{1KK}}{M_p}} \left(e^{-A_1}\right)^2\propto 10^{-9}\,,\qquad 
\frac{H_{final}^{BU}}{\e_{2c}^{1/4}}\sim 10^{-16}\,,
\eqlabel{gaugeec}
\end{equation}
to an excellent approximation expanding KT plasma can be considered to be in  thermal equilibrium in 
(almost) flat space-time during the whole period to expansion, up to the confinement phase transition. 

For the Black Universe evolution dual to the KT plasma expansion (cooling) with the subsequent first order confinement phase 
transition, we expect that the energy density of the black hole in the long throat will dilute, according to the 
four-dimensional  radiation dominated cosmology. 
In higher dimensional picture this corresponds to the recession of the horizon in the direction towards imaginary   tip
of the throat. 
As the black hole horizon energy density redshifts below the 
critical energy density, a first order  phase transition must take place that would remove the horizon from 
the long throat and expose the tip of the throat\footnote{A 'horizon removal' transition
was discussed previously in \cite{cnr}. It is not clear that our transition is similar. We comment in the conclusion 
how one can study such transition in the context of gauge/string duality.}. 
The disappearance of the horizon will occur in patches in accordance with the dual picture of the 
first order phase transitions.

 Instead of the geometry with the horizon, the long throat will be filled up with the KK excitations.
If there were light SM fields, say, attached to probe brane(s) around the tip of the long throat,
they would be produced due to the  decay  these  KK modes.
It is at this stage that our universe is 'born', and  the SM hot FRW cosmology follows. 
In this case one could   estimate  the reheating temperature of the universe in the  scenario.
The energy density at the phase transition is $\e_{2c}\sim \left(100 { GeV}\right)^4$
Assuming that all this energy  is available for the reheating of the SM,
$\e_{2c}\sim \frac{\pi^2}{30}g_* T_{RH}^4$ (where $g_*\propto 10^2$ is the number of the Standard Model degrees of freedom),
 we find a relatively low  
reheating temperature $T_{RH} \sim 50$ GeV.

 However, this may be irrelevant, becase there is a new problem in the cosmological scenario with hierarchical throats.
Indeed,  it was assumed for sucessful phenomenology
that the probe  brane(s) containing SM field are located in the long throat from the very beginning.
Meanwhile, emergence of the Black Brane horizon cloaks the long throat together with any probe branes
located there. In a sense, the SM sector becomes screened from the theory. 
Later on, the geometrical phase transition with melting  horizon and re-appearance of the compact throat solution
is not accompanied  by the   re-appearance of the SM brane. The SM sector may be missing after the ``Black Universe'' epoch.

\section{Conclusion}
We presented a cosmological scenario when excited KK modes produced from the brane/antibrane annihilation 
in the inflationary throat tunnel to a Standard Model throat and cloak it with a Schwarzschild horizon.
This suggests a model when string theory inflation  is followed the 
`Black Universe' epoch. This is significantly different from the models of reheating
after string theory inflation previously considered in the literature. 
However, dual picture suggest that the `Black Universe' epoch can be  simply  described by the 
expanding  universe  filled  with the hot  plasma composed of 
light  particles of a (hidden sector) gauge theory 2 in deconfinement phase  with $K(T)$ degrees of freedom.
These   originate from the rapid decay of the massive particles of another, 
 confined (hidden sector) gauge theory 1.
After the first order phase transition, the gauge theory 2 is described by the confined phase.
If light SM particles are present in the theory, the 
 corresponding particles (glueballs) 
decay into  SM particles with relatively low reheat temperature $10-100 GeV$. 

However, for the hierarchical multiple throat scenarios
 we identify the problem of the SM sector disappearance: the probe brane with SM fields, which was initially
placed in the long throat, will be   absorbed  by the horizon
together with a  segment of the long throat. After the end of the  end ``Black Universe'' epoch and
re-appearnce of the long throat geometry, it is not clear how the probe could re-appear.
Similarly, it is not clear to us what is the dual gauge theory  interpretation
of the screening and re-appearance of the SM sector in the theory like (\ref{gauge}).

There are various directions for  further study. First, one has to resolve the problem of the  SM sector.
One potential resolution would be to keep SM brane far enough from the tip of the long throat 
so that is will be  not swallowed by the horizon.
On the cosmology side,  we expect a very drastic phenomena to occur at the epochs  of the black brane 'horizon emergence'
as well as  the 'horizon removal' transitions. 

'Horizon removal' phase transition is dual to the confinement/deconfinement phase transition 
of the long-throat dual gauge theory plasma, coupled to the 4d gravity. 
How one would study such a phase transition on the gravity side of the gauge/string correspondence?
In  \cite{b4} if was proposed how to study the gravity dual to a gauge theory in de-Sitter space-time. 
In our cosmological scenario the dual gauge theory plasma couples to an FRW cosmology, driven by the 
plasma energy density. Thus, extending ideas of \cite{b4}, we should try to set-up the boundary metric 
to that of the appropriate FRW cosmology. The full ten-dimensional geometry should then be reconstructed 
requiring the nonsingularity, along the line of the gravity dual to the boost-invariant 
expansion of the $\caln=4$ supersymmetric Yang-Mills plasma \cite{bi1,bi2,bi3,bi4}.

\section*{Acknowledgments}
We would like to thank Neil Barnaby, Renata Kallosh,  Igor Klebanov, Volodya Miransky, Rob Myers, Marco Peloso,
Joe Polchinski, Sergei Prokushkin
  and Dam Son   for valuable discussions.
We would like to thank KITP Santa Barbara for hospitality, where this work was initiated
during the Program 'Nonequilibrium Dynamics in Particle Physics and Cosmology''.
AB's research at Perimeter Institute is supported in part by the Government
of Canada through NSERC and by the Province of Ontario through MRI.
AB gratefully acknowledges further support by an NSERC Discovery
grant and support through the Early Researcher Award program by the
Province of Ontario. LK was supported by NSERC and CIFAR.

\end{document}